\documentclass[Afour,sagev,times]{sagej}

\usepackage{moreverb,url}
\usepackage[ruled]{algorithm2e}
\usepackage[colorlinks,bookmarksopen,bookmarksnumbered,citecolor=red,urlcolor=red]{hyperref}
\usepackage{graphicx} 

\newcommand\BibTeX{{\rmfamily B\kern-.05em \textsc{i\kern-.025em b}\kern-.08em
T\kern-.1667em\lower.7ex\hbox{E}\kern-.125emX}}

\def\volumeyear{2022}

\begin{document}

\runninghead{Liu et al.}

\title{Reliability Analysis of Complex Multi-State System Based on Universal Generating Function and Bayesian Network}

\author{Xu Liu\affilnum{1}, Wen Yao\affilnum{1}, Xiaohu Zheng\affilnum{1,2} and Yingchun Xu\affilnum{3}}

\affiliation{\affilnum{1}Defense Innovation Institute, Chinese Academy of Military Science, Beijing 100071, China\\
\affilnum{2}College of Aerospace Science and Engineering, National University of Defense Technology, Changsha 410073, China\\
\affilnum{3}Xichang Satellite Launch Center, No.3, North Road, Xichang, Sichuan province, China}

\corrauth{Wen Yao, Defense Innovation Institute, Chinese
Academy of Military Science, Beijing 100071, China.}

\email{wendy0782@126.com}

\begin{abstract}
In the complex multi-state system (MSS), reliability analysis is a significant research content, both for equipment design, manufacturing, usage and maintenance. Universal Generating Function (UGF) is an important method in the reliability analysis, which efficiently obtains the system reliability by a fast algebraic procedure. However, when structural relationships between subsystems or components are not clear or without explicit expressions, the UGF method is difficult to use or not applicable at all. Bayesian Network (BN) has a natural advantage in terms of uncertainty inference for the relationship without explicit expressions. For the number of components is extremely large, though, it has the defects of low efficiency. To overcome the respective defects of UGF and BN, a novel reliability analysis method called UGF-BN is proposed for the complex MSS. In the UGF-BN framework, the UGF method is firstly used to analyze the bottom components with a large number. Then probability distributions obtained are taken as the input of BN. Finally, the reliability of the complex MSS is modeled by the BN method. This proposed method improves the computational efficiency, especially for the MSS with the large number of bottom components. Besides, the aircraft reliability-based design optimization based on the UGF-BN method is further studied with budget constraints on mass, power, and cost. Finally, two cases are used to demonstrate and verify the proposed method.

\end{abstract}


\maketitle
\section{Introduction}
When the system becomes more complex, the system and its components have different states characterized by different levels of performance. Such a system is referred to as a multi-state system \cite{harney2006controlling} (MSS). The MSS has a finite number of performance rates (intensity of the task accomplishment), which performs its tasks with various performance rates. Failures of some components may lead to the degradation of the system performance. Thus the reliability analysis of the MSS has recently received substantial attention  \cite{xiao2016optimal, azadeh2015multi}. There are four main methods for reliability analysis of the MSS: stochastic process method \cite{nakamura2013stochastic}, Universal Generating Function (UGF) \cite{youssef2008performance, babaei2022universal}, Bayesian Network (BN) \cite{zheng2019improved} and Monte Carlo simulation \cite{patowary2018reliability}. Among these methods, in terms of efficiency, the UGF method is commonly used for the reliability analysis of the MSS.

UGF is a reliability analysis method that holds the bare-looking and easily program-realized merits for the MSS \cite{ li2012multi, wen2018reliability, zheng2022performance}. Besides, the UGF method extends the moment-generating function and reduces the computational complexity for reliability assessment of the MSS. Due to its efficiency, the UGF method efficiently analyzes the reliability of the system and is suitable for solving different reliability problems of the MSS \cite{bao2019multi, zhang2022reliability, babaei2022estimating}. The traditional UGF method, however, only solves scenarios where the state performance of a component as well as the corresponding probability are real values and the relationships between components are dependent. To calculate the reliability of the system with fuzzy values, the fuzzy UGF method is presented to extend the traditional UGF with crisp sets \cite{gao2018dynamic, ding2008fuzzy}. Based on the fuzzy UGF method, Dong et al. \cite{dong2019reliability} analyze a generalized standard uncertain number to uniformly represent multi-source heterogeneous uncertain data and Qiu et al. \cite{qiu2021fuzzy} evaluate the fuzzy reliability of series systems with performance sharing between adjacent units. Considering the dependency of functions and operations in a mission or a system, Levitin et al. \cite{levitin2004universal} propose the modified UGF method to analyze the information processing system with dependent elements, and Farsi et al. \cite{farsi2017develop} propose an another modified UGF method to model and assess reliability for a solar array mechanism. The current research about the UGF method mainly focuses on the description of fuzzy values and independence between components for the MSS without explicit expressions. When the system structure is extremely complex, namely, it is difficult to describe with explicit expressions, the traditional UGF method is hardly used to analyze the system reliability.

BN is a useful tool to calculate the system reliability, which has the comprehensive reasoning ability for the complex MSS \cite{pearl2014probabilistic, li2019stochastic}. For representing the complex MSS, Nielsen et al. \cite{nielsen2013using} develop the Reduced Ordered Binary Decision Diagram (ROBDD) to efficiently perform the reliability inference in BN, but the number of paths in the ROBDD is exponentially increasing as the number of components increases. Apparently, the number of state combinations of units will also increase exponentially. To solve this problem, Tien et al. \cite{tien2016algorithms} propose the compression algorithm and inference algorithm the reliability assessment of infrastructure systems. Based on the compression algorithm and inference algorithm, Zheng et al. \cite{zheng2019improved} propose the improved compression inference algorithm and extend it to the multi-state nodes with independent binary parent nodes. In theory, these methods converge to the exact solution with a sufficiently large number of samples \cite{geduk2021practical}. In practice, however, the rate of convergence is unknown and slow \cite{straub2009stochastic}. Thus, although the BN method has the advantages in uncertainties, the computational efficiency is low for the reliability analysis of the complex MSS.

Generally, structure-function relationships between bottom components are not complicated for most MSSs, while structure-function relationships between subsystems or components become more complex as the number of system levels increases. Therefore, UGF is used to analyze system levels with simple structure relationship and BN is utilized to solve system levels that structure relationships are without explicit expressions. To this end, this paper proposes a method called UGF-BN by combining the respective superiority of UGF and BN to improve the efficiency of reliability analysis. Concretely, the UGF method is used to describe state performances of bottom components so that probability distributions of bottom components can be obtained efficiently. Then, for the structural relationships without explicit expressions, the BN method is used to analyze the system reliability based on the probability distributions obtained of bottom components. Besides, the aircraft reliability-based design optimization based on the UGF-BN method is further studied with budget constraints on mass, power, and cost. The main contributions of this paper include two parts: (1) the UGF-BN method is considerably cheaper in terms of computational cost than the BN method. (2) the UGF-BN method is further applied to efficiently construct the aircraft reliability-based design optimization model.

The remainder of this paper is structured as follows. In section ``Background: A brief introduction to Universal Generating Function and Bayesian Network", a brief background about UGF and BN is provided wherein the basic conception and the respective characteristics are presented. To combine the respective superiority, a novel method called UGF-BN is proposed to calculate the reliability of the complex MSS in section ``The UGF-BN method". Besides, the UGF-BN method is further studied for the aircraft reliability design optimize with budge mass, cost, power, and cost in section ``Aircraft reliability-based design optimization". Then two cases are studied to demonstrate and verify the proposed method in section ``Case study". The conclusion and future work are summarized in section ``Conclusion".

\vspace{20 pt}
\vspace{12 pt}

\section{Background: A brief introduction to Universal Generating Function and Bayesian Network}\label{sec2}

\subsection{The Universal Generating Function method}
UGF is an important reliability analysis technique that holds the bare-looking and easily program-realized merits for the complex MSS. The traditional UGF method adopts a fast algebraic procedure to obtain system reliability efficiently. However, when the structural relationship between subsystems or components is not clear or it is difficult to describe with explicit expressions, the traditional UGF method is difficult to implement or not applicable at all. 

The traditional UGF method are described in detail below. Suppose that a discrete random variable $G$ has probability distributions characterized by the vector $g$ and the vector $p$, which can be formulated by

\begin{equation}
\label{eq1}
\left\{ \begin{array}{l}
g=\left( g_1,\cdots ,g_K \right),\\
p=\left( p_1,\cdots ,p_K \right),\\
p_{i}=Pr\left( G=g_i \right).\\
\end{array} \right.
\end{equation}
According to the basic principle of the UGF method, the u-function of the independent variable $G$ is defined as
\begin{equation}
\label{eq2}
u_{j}(z)=\sum_{i=1}^{K_{j}} p_{j,i} z^{g_{j,i}},
\end{equation}
where the variable $G$ has $K_{j}$ possible values, $p_{j,i}$ represents the probability when the variable $G$ is equal to $g_{j,i}$, the variable $z$ is used to provide polynomial-like formulas, and the performance information is stored in the exponent.

To obtain the u-function of the subsystem that contains two components, the composition operator is introduced. For the system, its u-function is calculated as
\begin{equation}
\label{eq3}
\begin{aligned} u_{j}(z) \underset{w}{\otimes}u_{l}(z) &=\sum_{i_{j}=1}^{K_{j}} p_{j,i_{j}} z^{g_{j,i_{j}}} \underset{w}{\otimes} \sum_{i_{l}=1}^{K_{l}} p_{l,i_{l}} z^{g_{l,i_{l}}} \\ &=\sum_{i_{j}=1}^{K_{j}} \sum_{i_{l}=1}^{K_{l}} p_{j,i_{j}} p_{l,i_{l}} z^{w\left(g_{j,i_{j}}, g_{l,i_{l}}\right)},
\end{aligned}
\end{equation}
where the function $w(\cdot)$ is strictly defined by the type of connection between components on the structure of the logic-diagram representing the subsystem. The system structure-function $w(\cdot)$ should be substituted by $par(\cdot)$ and $ser(\cdot)$ in accordance with the type of connection between the components. The composition operators $par(\cdot)$ and $se r(\cdot)$ are defined for the parallel and series connections of a pair of components, but it can not represent a situation where the connection is unclear or without explicit expressions.

For a system containing $n$ components, a general composition operator ${\Psi_{w}}$ is used to represent the system. Its u-function is computed as

\begin{equation}
\label{eq4}
\begin{aligned} U(z)&=\Psi_{w}\left(u_{1}, \cdots, u_{n}\right) \\&=\sum_{i_{1}}^{K_{1}} \cdots \sum_{i_{1}}^{K_{n}} \prod_{j=1}^{n} p_{j i_{j}} z^{w\left(g_{1 i}, g_{2 i_{2}}, \cdots, g_{n i_{n}}\right)}.\end{aligned}
\end{equation}

\begin{figure}[!htbp]
	\centering
	\includegraphics[scale=1]{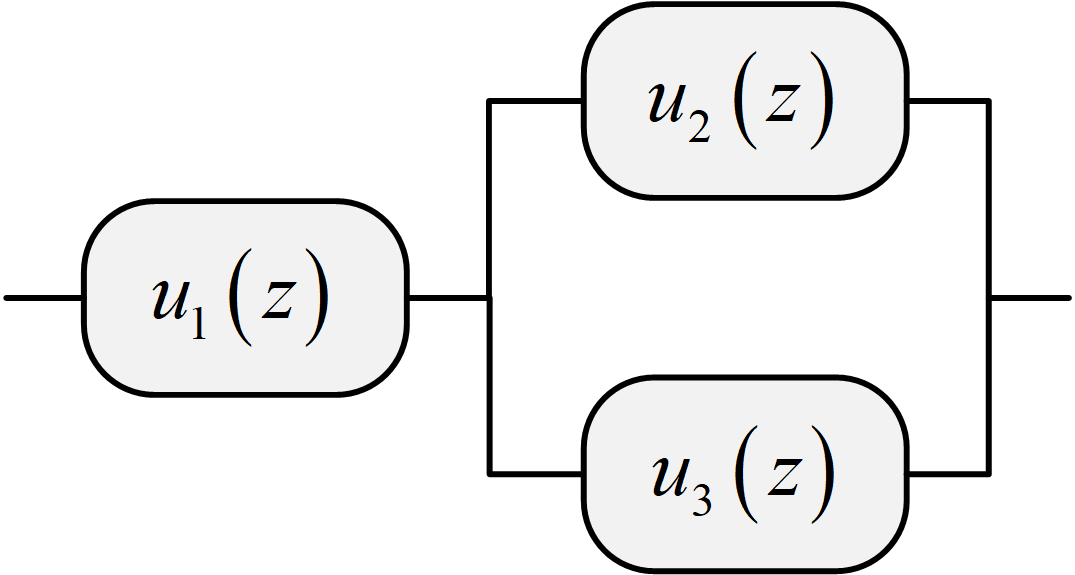}
	\caption{An aircraft information transmission system structure.}
	\label{UGF}
\end{figure}

For illustration, an aircraft information transmission system is shown in Figure \ref{UGF}, which consists of three components. The component 1 has two states $\{0,1\}$ with the performance rates $\{g_{11}=0,g_{12}=1\}$, and the corresponding probabilities $\{p_{11}=0.4,p_{12}=0.6\}$. The component $2$ has three states $\{0,1,2\}$ with the performance rates $\{g_{21}=0,g_{22}=1,g_{23}=2\}$, and the corresponding probabilities $\{p_{21}=0.2,p_{22}=0.3,p_{23}=0.5\}$. The component $3$ has two states  $\{0,1\}$  with the performance rates $\{g_{31}=0,g_{32}=1\}$, and the corresponding probabilities $\{p_{31}=0.5,p_{32}=0.5\}$. The system is connected in parallel-series, therefore the system structure-function is denoted as $w=\min \left(G_{1}, G_{2}+G_{3}\right),$ where $G_{1} \in\left\{g_{11}, g_{12}\right\}, G_{2} \in\left\{g_{21}, g_{22}, g_{23}\right\}$ and $G_{3} \in\left\{g_{31}, g_{32}\right\}$. According to Eq.(\ref{eq2}), the u-function for each component is defined as

\begin{equation}
\label{eq5}
\left\{\begin{array}{l}{u_{1}=0.4 z^{0}+0.6 z^{1}}, \\ {u_{2}=0.2 z^{0}+0.3 z^{1}+0.5 z^{2}}, \\ {u_{3}=0.5 z^{0}+0.5 z^{1}}.\end{array}\right.
\end{equation}
Applying the system structure-function $\Psi_{w}$, the u-function of the entire system is calculated as
\begin{equation}
\label{eq6}
\begin{aligned} 
U(z) &=\Psi_{w}\left(u_{1}, u_{2}, u_{3}\right) \\ 
					 &=0.46 z^{0}+0.54 z^{1} .
\end{aligned}
\end{equation}

\subsection{The Bayesian Network method}
The BN method is a vital tool for the reliability analysis, which has the comprehensive reasoning ability and natural advantages for uncertainty inference. The BN method uses a directed acyclic graph combining graph theory and probability theory. On the one hand, the graphical language of the graph theory is used to visually describe structure relationships of subsystems. On the other hand, the solution of uncertainty inference can be efficiently processed according to the probability theory. However, once the number of components in the MSS is extremely large, the computational cost of reliability analysis will increase exponentially. 

\begin{figure}[!htbp]
	\centering
	\includegraphics[scale=1]{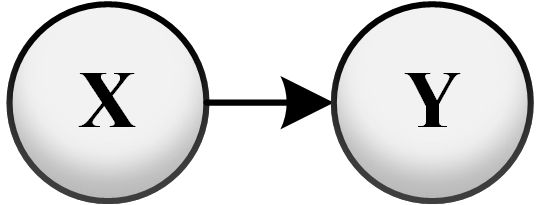}
	\caption{Digraph from $X$ to $Y$.}
	\label{BN}
\end{figure}

The BN method is described in detail below. A BN is characterized by a directed acyclic graph combining the graph theory and the probability theory. Some basic concepts in graph theory are introduced. As shown in Figure \ref{BN}, if there is a directed arc from node $X$ to node $Y$, the node $X$ is the parent node, and the node $Y$ is the child node of $X$. If there is no parent node, the node $X$ is called the root node.

In a directed acyclic graph, the node represents the random variable, and the directed arc between nodes represents the dependence relationship between two random variables. Each node has a Condition Probability Table (CPT). The CPT of root nodes is the marginal probability distribution, while the CPT of non-root nodes is the conditional probability distribution. The basic definition of a BN is as follows. A BN includes a binary group $X=\left(G,P\right)$, and $G=\left(V,A\right)$ is a directed acyclic graph. $P=\left(P\left(V_{i}\right)|\mu(V)| i=1,2, \cdots, n\right)$ is the CPT set of all nodes in the directed acyclic graph, where $n$ denotes the number of all nodes, $V_{i}$ denotes the $i\,\text{th}$ node, $V$ denotes a set of all nodes, $A$ denotes the set of directed arcs, and $\mu(V_{i})$ denotes the set of the parent of $V_{i}$. If $V_{i}$ is the root node, $\mu\left(V_{i}\right)=\varnothing$.

\begin{figure}[!htbp]
	\centering
	\includegraphics[scale=1]{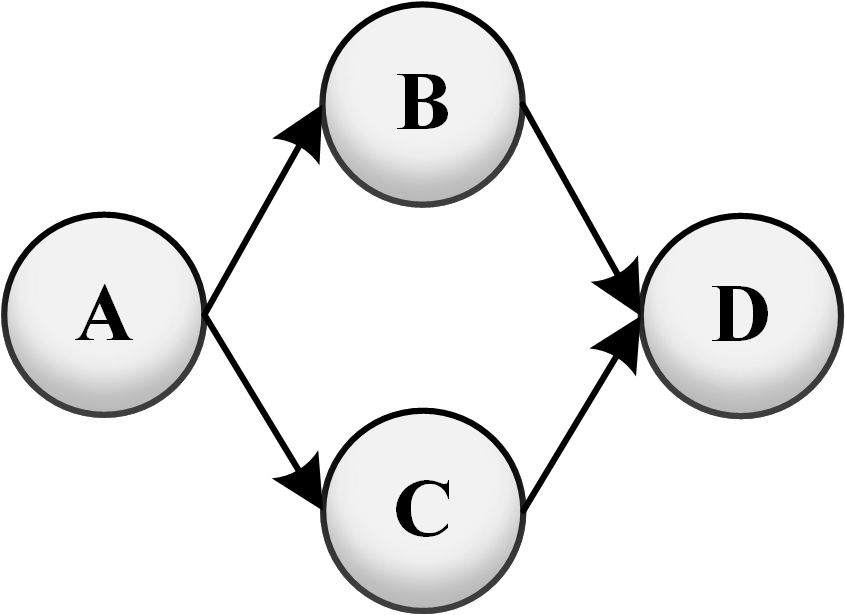}
	\caption{A simple case of the BN.}
	\label{BN2}
\end{figure}

For illustration, as shown in Figure \ref{BN2}, it is a simple BN node. $A$ is the parent node of $B$ and $C$, and $D$ is the child node of $B$ and $C$. Each node has an independent CPT. The probability distribution of $A$ is listed in Table \ref{tab1}. The nodes $A$ and $B$ have two states $\{0,1\}$. According to the relationship between nodes $A$ and $B$, the CPT of node $B$ is obtained in Table \ref{tab2}, thus the BN inference is realized. The probability distribution of $C$ and $D$ can be obtained in the same way.

\begin{table}[!htbp]%
	\centering
	\caption{Probability distribution of the node $A$.\label{tab1}}%
	\scalebox{0.88}{
	\begin{tabular*}{240pt}{@{\extracolsep\fill}ccc@{\extracolsep\fill}}%
		\toprule
		\textbf{state} & $P(A)=0$ & $P(B)=1$ \\
		\midrule
		probability value & 0.5 & 0.5 \\
		\bottomrule
	\end{tabular*}
	}
\end{table}

\begin{table}[!htbp]%
	\centering
	\caption{The CPT of node $B$.\label{tab4}}%
	\scalebox{0.88}{
	\begin{tabular*}{240pt}{@{\extracolsep\fill}ccc@{\extracolsep\fill}}%
		\toprule
		\textbf{state of $B$} & $P(A)=0$ & $P(A)=1$ \\
		\midrule
			0 & 0.2 & 0.8 \\
			1 & 0.7 & 0.3 \\
		\bottomrule
	\end{tabular*}
	}
	\label{tab2}
\end{table}



\begin{algorithm*}[!htbp]
	\caption{\textbf{Pseudo code of the UGF-BN algorithm}}
	\label{alg_UGF-BN}
	\LinesNumbered
	\KwIn{\\
	(1) Structure-functions for bottom components:\\
	$\left\{\Psi_{\omega_{L}} \mid 1 \leq L \leq m_{2}\right\}$\\
	(2) Relationships without explicit expressions in level-2 to level-n\\
	(3) Probability distribution of all bottom components in level-1.
	}
	\KwOut{\\Complex multi-state system reliability $R_{\text {system }}$.}
	Construct the hierarchical system for the whole MSS;\\
	Calculate u-function of each component in level-1;\\
	Calculate probability distributions of each node in level-2, given $\left\{\Psi_{\omega_{L}} \mid 1 \leq L \leq m_{2}\right\}$ in level-1; \\
	Take the probability distributions obtained as root nodes and construct BN model from level-2 to level-n;\\
	Do step 6 and step 7 for level-2 to level-n to obtain the probability distributions of the node in level-n;\\
	Establish the CPTs of child nodes according to the relationships between parent nodes and child nodes;\\
	Conduct BN inference to acquire probability distributions of child nodes;\\
	Obtain $R_{\text {system }}$, given the probability distributions obtained in level-n; 
\end{algorithm*}

\section{The UGF-BN method}\label{sec3}
\begin{figure*}[!htbp]
	\centering
	\includegraphics[scale=0.3]{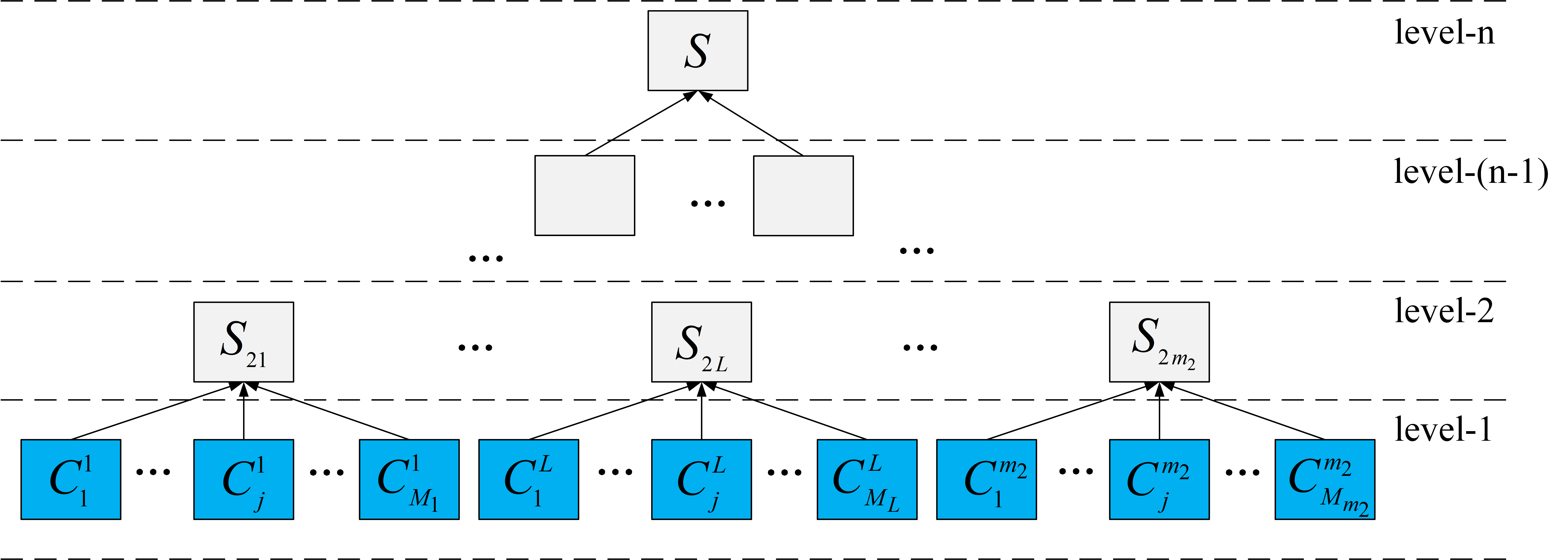}
	\caption{The construction of the hierarchical system.}
	\label{hi}
\end{figure*}

In most cases, such as the complex multi-state hierarchical system in Figure \ref{hi}, the structure-function relationships between bottom components are not complex, but the number of bottom components is considerably large. Besides, the structure-function relationships of subsystems or systems become more complex as the number of system levels increases. For the hierarchical system, using the BN method to analyze the whole system requires large computational costs for the reliability analysis. If the UGF method is adopted to analyze the whole system, since the structure-function relationships of the system are without explicit expressions, it is difficult and even intractable to obtain the system reliability accurately. 

\begin{figure}[!htbp]
	\centering
	\includegraphics[scale=0.42]{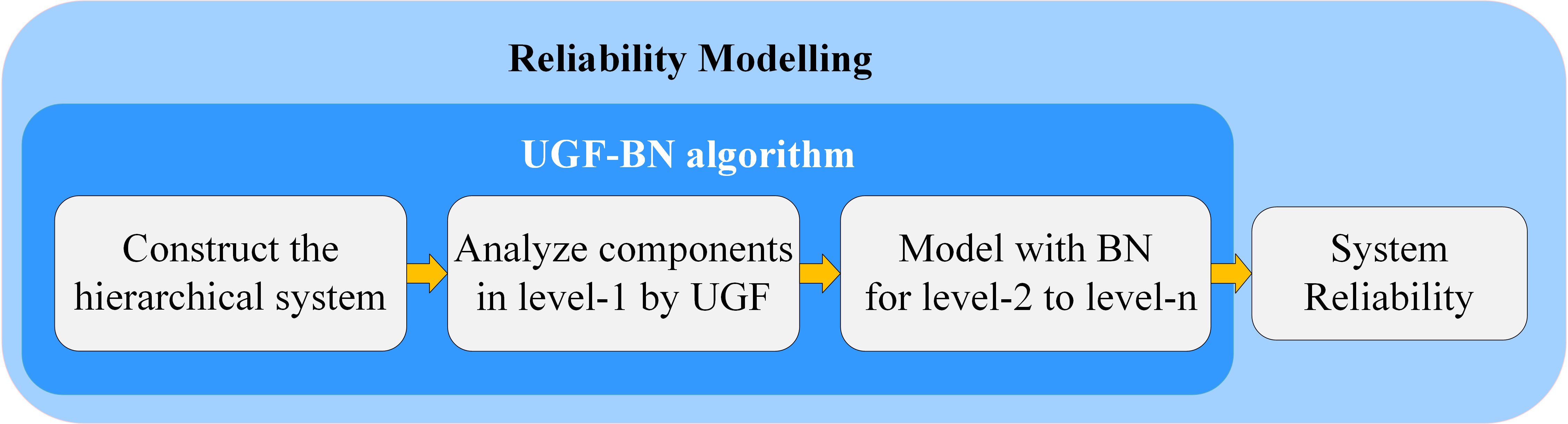}
	\caption{Main strategies of the UGF-BN method.}
	\label{Main}
\end{figure}

To this end, we propose the UGF-BN for reliability
analysis of the complex MSS, which consists of three main aspects as shown in Figure \ref{Main}, construction of the hierarchical system, acquisition of probability distributions of nodes in level-2 by UGF, and reliability analysis by BN. Firstly, the hierarchical system is constructed to describe the relationship from level-1 to level-n. Secondly, the u-function of each node in level-1 is acquired, then probability distributions of each node in level-2 are obtained through the system structure-function, and then probability distributions obtained are regarded as the input nodes in level-2. Finally, the BN model from level-2 to level-n is constructed, and then the CPT of child nodes is established according to the relationships between parent nodes and child nodes. The BN inference is conducted to acquire the probability distributions of the node in level-n. The MSS reliability $R_{\text {system }}$ is obtained through the probability distributions in level-n. In summary, the detailed pseudo-code of the UGF-BN method is shown in Algorithm \ref{alg_UGF-BN}. It is noteworthy that the UGF-BN method in this paper only uses UGF for level-1 and then constructs the BN model for level-2 to level-n, which just provides an idea. Spirited by this, the UGF method can be applied to different levels according to different strategies, and then the BN model is constructed, which will further improves the computational cost of the whole system.

\subsection{Construction of the hierarchical system}\label{sec3.1}
The hierarchical system is ubiquitous in engineering systems, and its obvious characteristic is that the output of low-level components is the input of high-level subsystems. Taking a complex MSS as shown in Figure \ref{hi} for example, the level is numbered in an ascending order from the bottom level to the top level, the bottom level is level-1 and the top level is level-n. Each node in level-1 is denoted by $C_{j}^{L}\left(0 \leq j \leq m_{L}\right)$, which represents the $j\,\text{th}$ component of the $L\,\text{th}$ node of level-2. The $L\,\text{th}$ node in $\text{level}-i$ $\left(0 \leq L \leq m_{i},2 \leq i \leq n\right)$ is represented by $S_{iL}$. The relationship between the components in level-1 is described by $\Psi_{w_{L}}\left(1 \leq L \leq m_{2} \right)$, where $\Psi_{L}$ represents the system structure-function of the $L\,\text{th}$ node of level-2.

\subsection{Acquisition of probability distributions of nodes in level-2 by UGF}\label{sec3.2}
For the $\text { L th }$ node in level-2, the node $S_{2 L}$ includes $M_{L}$ components, the probability distribution is represented by $g_{j_{j}}=\left(g_{j 1}, \cdots, g_{j K_{j}}\right)$, $p_{j_{j}}=\left(p_{j 1}, \cdots, p_{j K_{j}}\right)$ $\left(1 \leq j \leq M_{L}\right)$. The u-function of $M_{L}$ components is formulated as
\begin{equation}
\label{eq7}
u_{j}^{L}(z)=\sum_{i_{j}=1}^{K_{j}} p_{j,i_{j}} z^{p_{j,i_{j}}}\left(1 \leq j \leq M_{L}\right).
\end{equation}

Then, the probability distribution of the node $S_{2 L}$ in level-2 is obtained by the structure-function $\Psi_{w_{L}}$. The u-function of the $\text { L th }$ node in level-2 is computed by
\begin{equation}
\label{eq8}
\begin{aligned}
\mathrm{U}_{L}(\mathrm{z}) &=\Psi_{w_{L}}\left(\mathrm{u}_{1}^{L}, \cdots \mathrm{u}_{M_{L}}^{L}\right) \\
&=\sum_{i_{1}=1}^{K_{1}} \cdots \sum_{i_{M_{L}}=1}^{K_{M_{L}}} \prod_{j=1}^{M_{L}} \mathrm{p}_{j,i_{j}} \mathrm{z}^{w\left(\mathrm{g}_{1,i_{1}},\cdots,\mathrm{g}_{M_{L},i_{M_{L}}}\right)}.
\end{aligned}
\end{equation}

Similarly, the u-function of other nodes in level-2 is calculated in the same way, the probability distributions of each node in level-2 are represented by $U_{1}(z),\dots,U_{m_{2}}(z)$.

\subsection{Reliability analysis by BN}\label{sec3.3}
\begin{figure}[!htbp]
	\centering
	\includegraphics[scale=0.5]{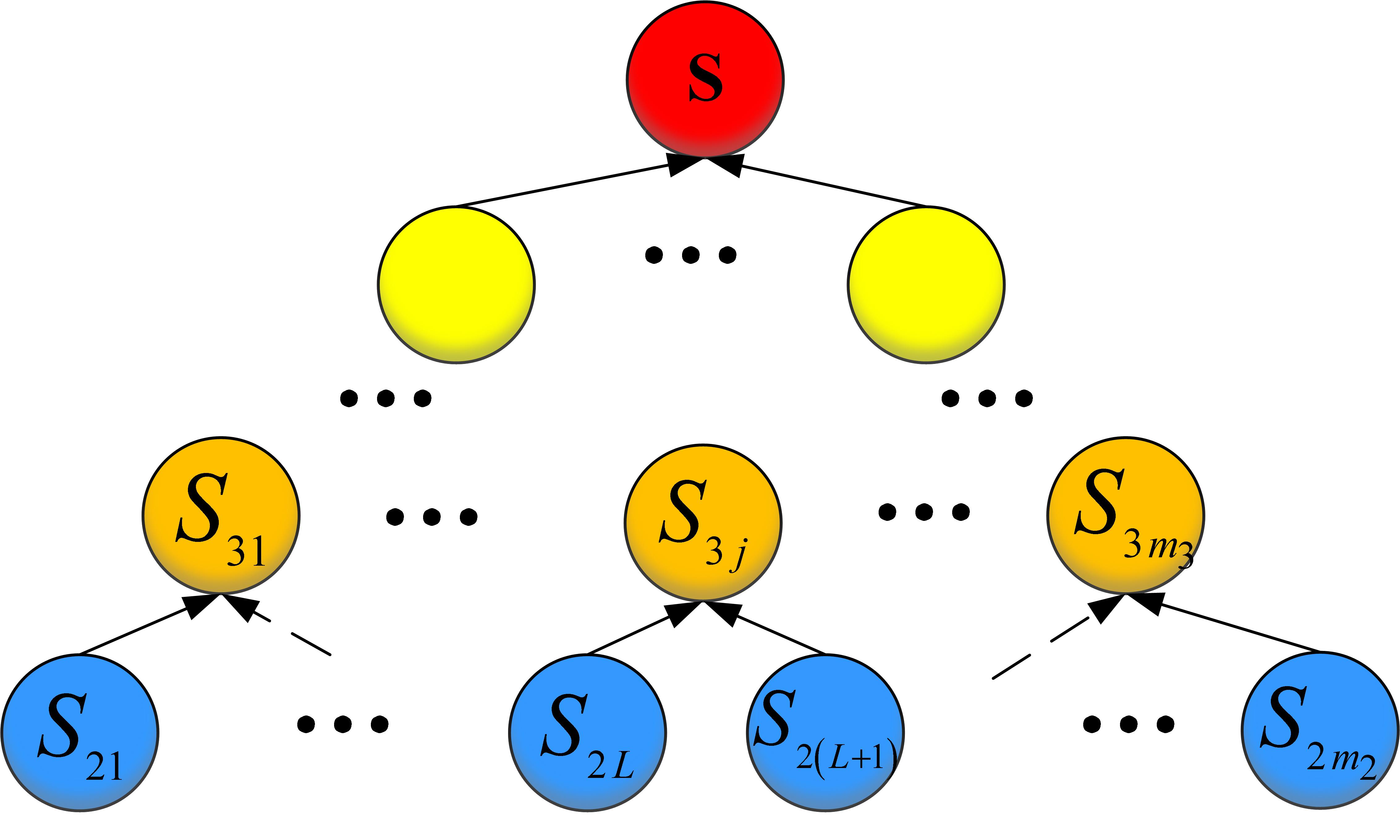}
	\caption{The construction of the hierarchical system.}
	\label{BN_hi}
\end{figure}

Firstly, the probability distributions obtained of each node in level-2 are taken as the probability distributions of root nodes of BN. According to the hierarchical system shown in Figure \ref{hi}, the BN model for the nodes from level-2 to level-n is constructed as shown in Figure \ref{BN_hi}.

For the BN model in Figure \ref{BN_hi}, taking the $\text { L th } $ node in level-2 for example, node $S_{3 j}$ is the child node of node $S_{2 L}$. Suppose that child node $S_{3 j}$ with three states has two parent nodes $S_{2 L}$ and $S_{2(L+1)}$. The node $S_{3 j}$ has two parent nodes $S_{2 L}$ and $S_{2(L+1)}$ and each parent node has three states. Thus, three nodes $S_{2 L}$, $S_{2(L+1)}$ and $S_{3 j}$ have the state set $\{0,1,2\}$, where $0$, $1$ and $2$ denote complete failure, degradation, and normal working respectively. According to relationships between parent nodes $S_{2 L}$ and $S_{2(L+1)}$, the CPT of the child node $S_{3 j}$ is established in Table \ref{tab3}, where $P_{0}$ denotes $P\left(S_{3 j}=0 \mid S_{2 L}, S_{2(L+1)}\right)$, $P_{1}$ denotes $P\left(S_{3 j}=1 \mid S_{2 L}, S_{2(L+1)}\right)$, and $P_{2}$ denotes $P\left(S_{3 j}=2 \mid S_{2 L}, S_{2(L+1)}\right)$. Then the BN inference is conducted to acquire probability distributions of the child node in level-3.




Finally, similarly, the BN inference is conducted from level-3 to level-n to obtain probability distributions. Based on the probability distributions in level-n, the complex MSS reliability $R_{system}$ can be obtained.


\begin{center}
	\begin{table}[!htbp]%
		\centering
		\caption{The CPT of child node $S_{3 j}$.\label{tab3}}%
			\scalebox{0.88}{
		\begin{tabular*}{250pt}{@{\extracolsep\fill}ccc@{\extracolsep\fill}ccc@{\extracolsep\fill}ccc@{\extracolsep\fill}}%
			\toprule
			$S_{2L}$ & $S_{2(L+1)}$ & $P_{0}$ & $P_{1}$ & $P_{2}$ \\
			\midrule
			0 & 0 & $p_{11}$ & $p_{12}$ & $p_{13}$   \\
			0 & 1 & $p_{21}$ & $p_{22}$ & $p_{23}$  \\			
			0 & 2 & $p_{31}$ & $p_{32}$ & $p_{33}$   \\
			1 & 0 & $p_{41}$ & $p_{42}$ & $p_{43}$  \\	
			1 & 1 & $p_{51}$ & $p_{52}$ & $p_{53}$   \\
			1 & 2 & $p_{61}$ & $p_{62}$ & $p_{63}$  \\			
			2 & 0 & $p_{71}$ & $p_{72}$ & $p_{73}$  \\
			2 & 1 & $p_{81}$ & $p_{82}$ & $p_{83}$ \\	
			2 & 2 & $p_{91}$ & $p_{92}$ & $p_{93}$  \\	
			\bottomrule
		\end{tabular*}
		}
	\end{table}
\end{center}

\vspace{12 pt}
\vspace{16 pt}

\section{Aircraft reliability-based design optimization}\label{sec4}
\subsection{Aircraft optimization modeling}
Reliability is an important part of reliability-based design optimization. Based on the UGF-BN method, the aircraft reliability-based design optimization is further researched in this section. The reliability-based design optimization problem for the aircraft system design includes design variable and constraint conditions, which is formulated as follows.
\begin{itemize}
	\item [1)] Design variable.
\end{itemize}	

In the paper, the design variables are the number of bottom components $C_{j}$ in level-1 (omit superscripts for simplicity in this case) that selected for the aircraft. Taking the hierarchical system in the Figure \ref{hi} for example, the number of different components in level-1 is denoted as $n_{j}$ and $\left\{n_{1}, n_{2}, \cdots,\quad n_{\text {total }}\right\}$ are the design variables. Due to the design requirements of the aircraft, there are the maximum and minimum bounds of $C_{j}$, which are denoted as $n_{j}^{\max }$ and $n_{j}^{\min }$. In the reliability-based design optimization design space, the design variable is defined as

\begin{equation}
n_{j}^{\min } \leq n_{j} \leq n_{j}^{\max}.
 \end{equation}

\begin{itemize}
	\item [2)] Constraint conditions
\end{itemize}

In the overall design process of the aircraft, the main constraints to be considered for the aircraft reliability optimization are total mass, power, and cost, which are discussed separately as follows.

\textcircled{1} Mass constraint

Given the design variable $\left\{n_{1}, n_{2}, \cdots, n_{\text {total }}\right\}$, the overall mass $M_{sum}$ of the system is calculated as
\begin{equation}
M_{sum}=\sum_{\left\{n_{1}, n_{2}, \cdots, n_{\text {total }}\right\}} m_{j} n_{j},
\label{eq_sum_mass}
\end{equation}
where $m_{j}$ is the mass of the $\text { j th }$ component in level-1. The overall mass $M_{sum}$ should be less than the budget $M_{budget}$ of the aircraft, i.e.,
\begin{equation}
M_{sum} \leq M_{budget}.
\label{eq_budget_mass}
\end{equation}

\textcircled{2} Power constraint

Given the design variable $\left\{n_{1}, n_{2}, \cdots, n_{\text {total }}\right\}$, the overall power $P o_{sum}$ of the system is calculated as follows:

\begin{equation}
\label{eq_sum_pow}
Po_{sum}=\sum_{\left\{n_{1}, n_{2}, \cdots, n_{\text {total }}\right\}} Po_{j} \times n_{j},
\end{equation}
where $Po_{j}$ is the power of the $\text { j th }$ component of the level-1. The overall power $Po_{sum}$ should be no more than the budget power $Po_{budget}$ of the aircraft, i.e.,

\begin{equation}
\label{eq_budget_pow}
Po_{sum}\leq Po_{budget}.
\end{equation}

\textcircled{3} Development cost constraint

Given the design variable $\left\{n_{1}, n_{2}, \cdots, n_{\text {total }}\right\}$, the overall development cost $Dc_{sum}$ of the aircraft is calculated as follows:

\begin{equation}
\label{eq_sum_dc}
Dc_{sum}=\sum_{\left\{n_{1}, n_{2}, \cdots, n_{\text {total }}\right\}} Dc_{j} n_{j},
\end{equation}
where $Dc_{j}$ is the development cost of the $j\,\text{th}$ component of the level-1. The overall power $Dc_{sum}$ should be no larger than the budget development cost $Dc_{budget}$ of the aircraft, i.e.

\begin{equation}
\label{eq_budget_bc}
Dc_{sum}\leq Dc_{budget}.
\end{equation}

\textcircled{4} Optimization objective

Reliability is significant for the aircraft design, use and maintenance. In this paper, the maximization of the reliability is chosen as the objective, which is:

\begin{equation}
\label{eq_max_r}
\text{max}R_{system}.
\end{equation}

\subsection{Aircraft reliability optimization formulation}
In the process of the reliability-based design optimization, given the design variable $\left\{n_{1}, n_{2}, \cdots, n_{\text {total }}\right\}$, the overall mass $M_{sum}$, the overall power $P o_{sum}$ and the overall development cost $D c_{sum}$ are calculated, respectively. Next, at the end of the aircraft lifetime, i.e., t = Life, the failure probability of each optional component in level-1 is calculated by its lifetime distribution. Finally, the aircraft reliability at the end of the lifetime is calculated by the UGF-BN method. The optimal design variables will be searched through the Sequence Quadratic Program algorithm \cite{huang2013sequential, luo2011modified, larson1992space} until the stopping condition is satisfied. To sum up, the satellite lifetime optimization problem is constructed as
\begin{figure}[!htbp]
	\centering
	\includegraphics[scale=0.65]{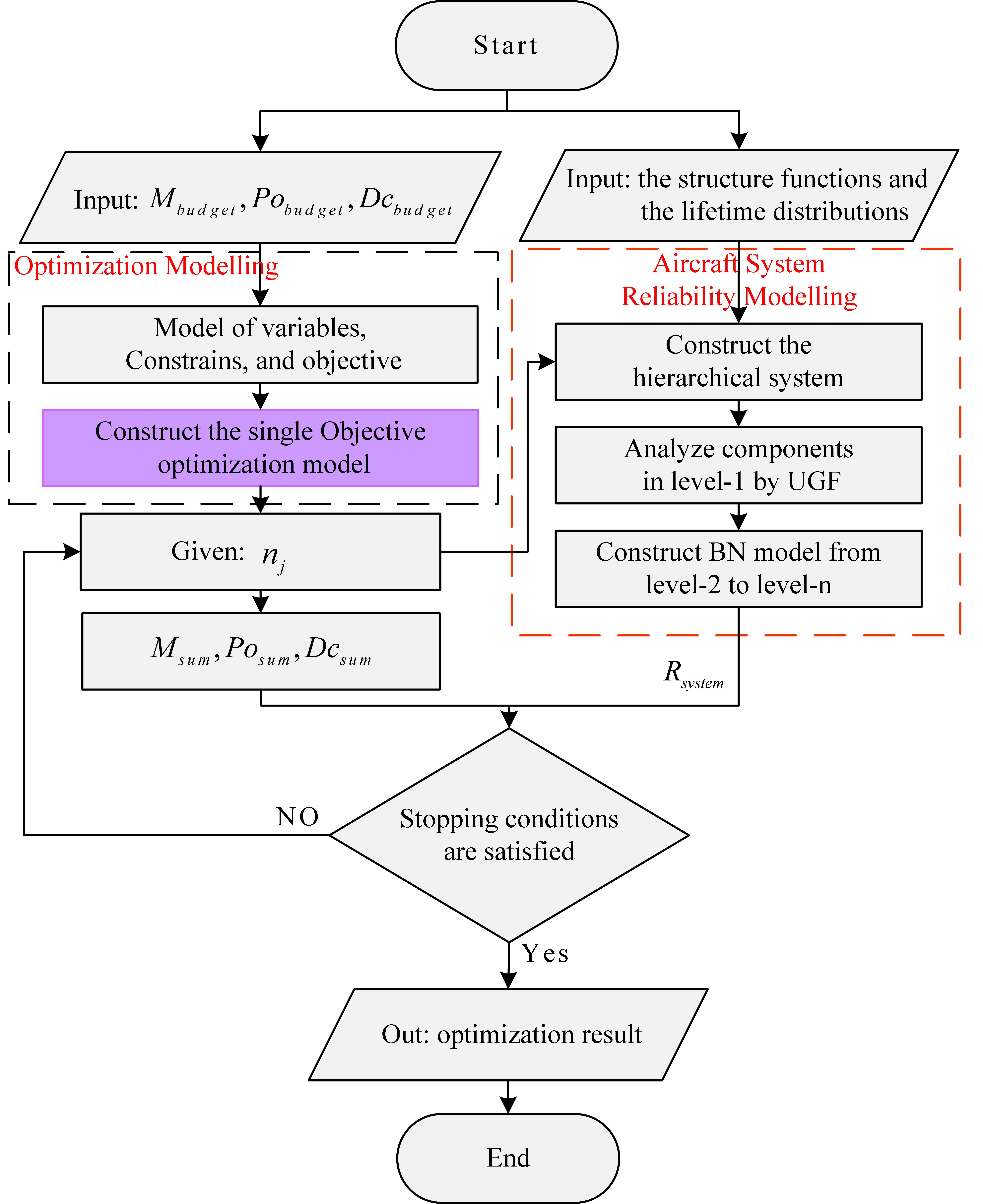}
	\caption{The flowchart of the aircraft reliability optimization process.}
	\label{flowchart}
\end{figure}

\begin{equation}
\label{eq20}
\left\{\begin{array}{l}
{\text { find } n_{j} \quad(j=1, \cdots, n_{\text {total}})} \\
{\text { max } R_{\text {system}}}, \\
{\text { s.t. } M_{sum} \leq M_{budget}}, \\
{P_{sum} \leq P_{budget}}, \\
{D c_{sum} \leq D c_{budget}}, \\
{R_{sum} \geq R_{budget}}, \\
{n_{j}^{\text {min}} \leq n_{j} \leq n_{j}^{\text {max}}.}
\end{array}\right.
\end{equation}
Based on the Sequence Quadratic Program algorithm, optimal design variables will be searched. The flowchart of the aircraft lifetime optimization process is shown in Figure \ref{flowchart}, and the main steps are summarized as follows:

\textbf{Step 1}: Determine the budgets of the total mass, the total power and the total cost for the aircraft to be designed. Besides, determine the structure-functions and the lifetime distributions of the bottom components of the aircraft system for the multi-state aircraft system reliability modeling.

\textbf{Step 2}: Construct the optimization modeling process. Do \textbf{step 2.1} and \textbf{step 2.2}.

\textbf{Step 2.1}: Determine the design variables, constraints and optimization objective.

\textbf{Step 2.2}: Construct the single objective optimization model.

\textbf{Step 3}: \textbf{The process of finding the optimal design variables}. Do \textbf{step 3.1}, \textbf{step 3.2} and \textbf{step 3.3}

\textbf{Step 3.1}: Give a scheme of the design variables as $\left\{n_{1}, n_{2}, \cdots n_{\text {total }}\right\}$.

\textbf{Step 3.2}: Calculate the total mass, power and cost by Eq.(\ref{eq_budget_mass}), Eq.(\ref{eq_budget_pow}) and Eq.(\ref{eq_budget_bc}). Meanwhile, based on the \textbf{UGF-BN method}, the reliability of the complex multi-state aircraft system can be obtained.

\textbf{Step 3.3}: Determine if the stopping condition is satisfied, and if so, do \textbf{step 4}, otherwise, go back to \textbf{step 3}.

\textbf{Step 4}: Output the final optimization result, and end the process.

\section{Case study}\label{sec5}
Two cases in this section are presented for demonstrating and verifying the proposed method. Case 1 utilizes an unmanned aircraft transmission system structure to demonstrate the efficiency of the UGF-BN method. Case 2 demonstrates a reliability-based design optimization problem of an autopilot software PX4 with the UGF-BN method.

\begin{figure*}[!htbp]
	\centering
	\includegraphics[scale=0.3]{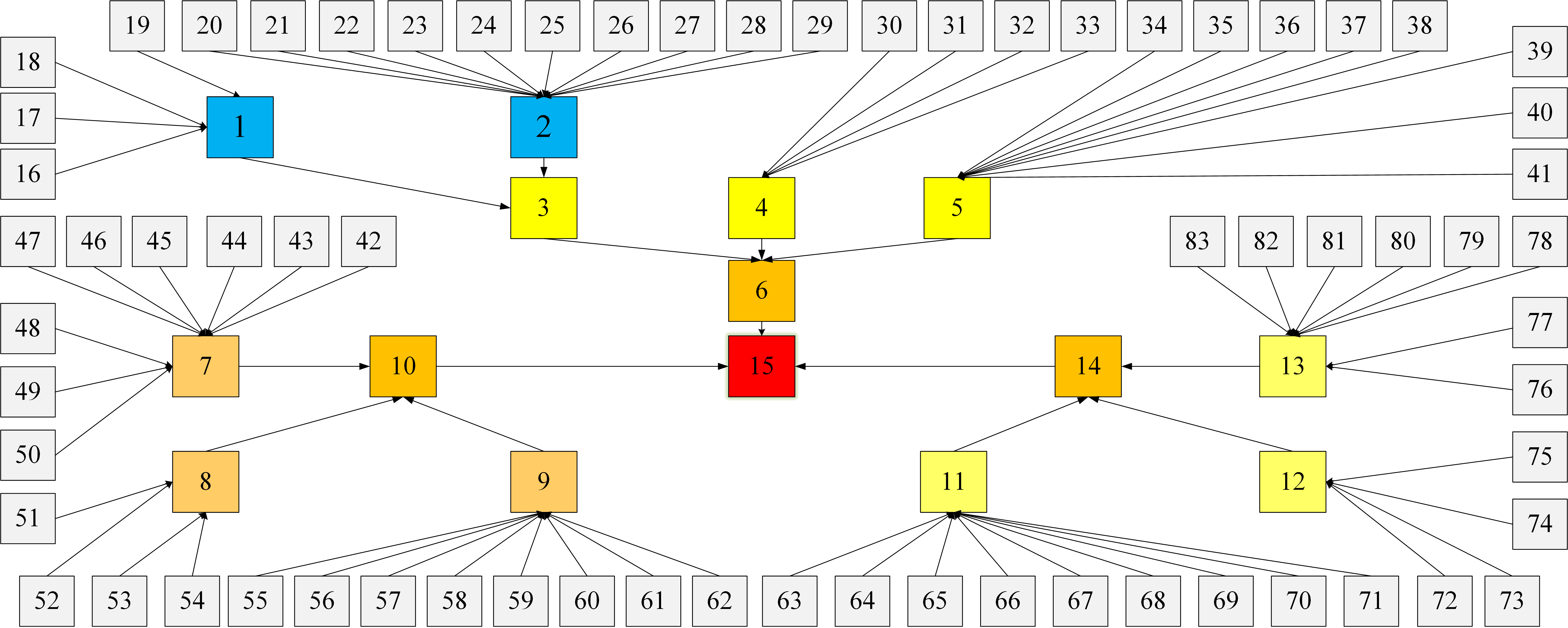}
	\caption{The hierarchical system of the unmanned aerial vehicle dynamical system.}
	\label{case1_hi}
\end{figure*}

\begin{table*}[!htbp]%
	\centering
	\caption{The descriptions of subsystems in the unmanned aerial vehicle dynamical system.\label{tab_subsystems_case1}}
		\scalebox{0.9}{
	\begin{tabular*}{530pt}{@{\extracolsep\fill}cc|c@{\extracolsep\fill}c|cc@{\extracolsep\fill}ccc@{\extracolsep\fill}ccc@{\extracolsep\fill}}%
		\toprule
		Number & Name of component & Number & Name of component & Number& Name of component \\
		\midrule
		1 & Conveyer driven shaft & 10,14 & Auxiliary power unit & 38,39& Tank   \\
		2 & Synchronizer gear & 15 & Power driven mechanism & 40,41 & Engine \\			
		3 & Synchronous gear drive mechanism & 16,17 & Hollow shaft & 42-45, 53-66 & Screw blade  \\
		4 & Main propeller & 18,19 & Central spindle & 46-50, 67-71 & Propeller blade attachment \\	
		5 & Power producer & 20-25 & Gear wheel & 51,52,72,73  & Motor \\
		6 & Main power group & 26-29 & Pinion & 53,54,74,75 & Motor accessories \\			
		7,11 & Auxiliary propeller & 30,31 & Positive rotation propeller & 55-62 & Battery pack \\
		8,12 & Electro-motor & 32,33 & Counter rotating propeller & 76-79 & Propeller accessories\\	
		9,13 & Battery pack & 34-37 & Tank accessories & 80-83 & Engine accessories \\	
		\bottomrule
	\end{tabular*}
	}
\end{table*}

\begin{table*}[!htbp]%
	\centering
	\caption{The probability distributions and relationships of bottom components in case 1.	\label{tab_probability_distribution_case1}}%
	\scalebox{0.93}{
	\begin{tabular*}{550pt}{@{\extracolsep\fill}ccc|@{\extracolsep\fill}ccc|@{\extracolsep\fill}ccc|@{\extracolsep\fill}ccc@{\extracolsep\fill}ccc@{\extracolsep\fill}ccc@{\extracolsep\fill}ccc@{\extracolsep\fill}}%
		\toprule
		Component & $\Psi_{i}$ & $\lambda(\times 10^{-6})$ & Component & $\Psi_{i}$& $\lambda(\times 10^{-6})$ & Component & $\Psi_{i}$& $\lambda(\times 10^{-6})$& Component & $\Psi_{i}$& $\lambda(\times 10^{-6})$ \\
		\midrule
		16 & 1 &  $1.52$&33 & 2 & $1.34$ &50 & 1 &  $1.22$&67 & 3 &  $9.86$ \\
		17 & 1 &$0.85$&34 & 1 &  $1.61 $&51 & 2 &  $1.61$&68 & 3 &  $1.52$\\
		18 & 1 &$0.93$&35 & 1 &  $1.55$&52 & 2 &  $1.55$&69 & 3 &  $1.73$\\
		19 & 1 &$0.98$&36 & 1 &  $1.16$&53 & 2 &  $1.16$&70 & 3 &  $2.13$\\
		20 & 3 & $1.55$&37 & 1 &  $1.21$&54 & 2 &  $1.71$&71 & 3 &  $9.51$\\
		21 & 3 & $1.72$&38 & 1 &  $1.12$&55 & 1 &  $1.52$&72 & 2 &  $9.98$\\
		22 & 3 & $2.11$&39 & 1 &  $1.32$&56 & 1 &  $1.42$&73 & 2 &  $2.14$\\
		23 & 3 & $0.96$&40 & 1 &  $1.47$&57 & 1 &  $1.22$&74 & 2 &  $1.14$\\
		24 & 3 & $1.00$&41 & 1 &  $1.73$&58 & 1 &  $1.63$&75 & 2 &  $1.63$\\
		25 & 3& $2.13$&42 & 1 &  $1.21$&59 & 1 &  $1.43$&76 & 1 &  $1.45$\\
		26 & 3& $2.24$&43 & 1 &  $0.99$&60 & 1 &  $1.51$&77 & 1 &  $1.36$\\
		27 & 3& $1.64 $&44 & 1 &  $1.53$&61 & 1 &  $1.72$&78 & 1 &  $1.61$\\
		28 & 3& $1.55 $&45 & 2 &  $1.73$&62 & 1 &  $1.32 $&79 & 1 &  $1.24$\\
		29 & 3& $1.26$&46 & 2 &  $2.13$&63 & 1 &  $1.41$&80 & 1 &  $1.12$\\
		30 & 2& $1.73$&47 & 2 &  $0.96$&64 & 3 &  $1.22$&81 & 1 &  $1.44$\\
		31 & 2& $1.32$&48 & 2 &  $1.00$&65 & 3 &  $1.52$&82 & 1 &  $1.63$\\
		32 & 2 & $1.12 $&49 & 1 &  $2.14$&66 & 3 &  $1.31$&83 & 1 &  $1.45$\\

		\bottomrule
	\end{tabular*}
	}
\end{table*}

\subsection{Case 1}
\subsubsection{Problem Description}
An unmanned aerial vehicle dynamical driven system has five levels, which has 15 subsystems, 68 bottom components, as is shown in Figure \ref{case1_hi}, where the gray represents bottom components and components with the same colors are represented in the same level. Besides, the detailed descriptions of subsystems are given in Table \ref{tab_subsystems_case1}. The probability distributions and relationships of bottom components are listed in Table \ref{tab_probability_distribution_case1}, where relationships include parallel, series and exclusive OR, which denote as $\{1,2,3\}$.

\subsubsection{System reliability modeling inference}

\begin{itemize}
	\item [1)] Construction of the hierarchical system
\end{itemize}

According to the system information that the system has five levels, which has 15 subsystems, 68 bottom components. The hierarchical system is constructed as shown in Figure \ref{case1_hi}. The lifetime probability functions of components in level-1 are exponential distributions. Through the parameters in Table \ref{tab_subsystems_case1}, the cumulative distribution function of the components in level-1 can be calculated as
\begin{equation}
F(t)=\int_{0}^{t} f(t) d t=\int_{0}^{t} \lambda e^{-\lambda t} d t=1-e^{-\lambda t}.
\end{equation}
Therefore, the probability distributions of the components in level-1 can be denoted by $\mathrm{F}(\mathrm{t})$ and $\mathrm{R}(\mathrm{t})$. And the reliability function can be obtained by
\begin{equation}
R(t)=1-F(t)=e^{-\lambda t}.
\end{equation}

\vspace{18 pt}

\begin{figure*}[!htbp]
	\centering
	\includegraphics[scale=0.45]{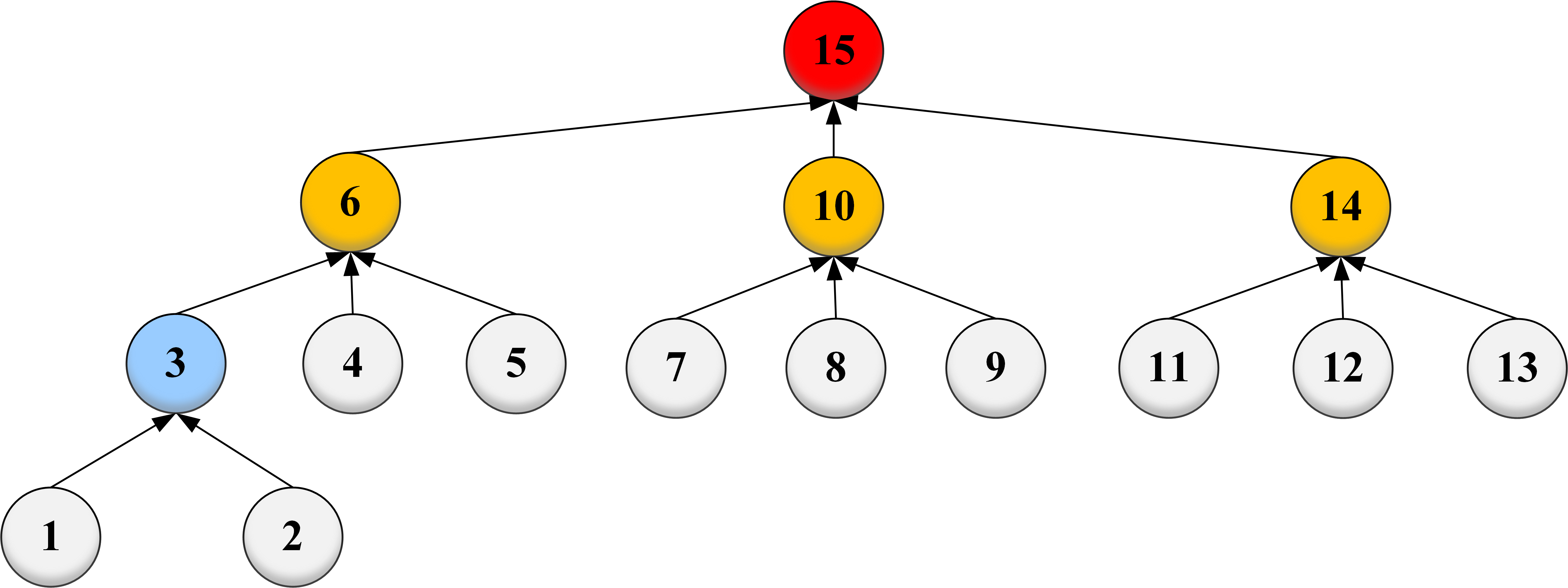}
	\caption{Multilevel BN model of the power driven system.}
	\label{case1_BN}
\end{figure*}

\begin{itemize}
	\item [2)] UGF modeling
\end{itemize}
For bottom components in level-1, the UGF method is firstly used to obtain the u-function. Taking the $\text{1th}$ node in level-2 for example, the node in level-2 includes four child nodes from node 16 to node 19. The u-function of the four child nodes in level-1 is described as
\begin{equation}
\left\{\begin{array}{l}
u_{16}^{1}=p_{16,1} z^{0}+p_{16,2} z^{1}, \\
u_{17}^{1}=p_{17,1} z^{0}+p_{17,2} z^{1}, \\
u_{18}^{1}=p_{18,1} z^{0}+p_{18,2} z^{1}, \\
u_{19}^{1}=p_{19,1} z^{0}+p_{19,2} z^{1}.
\end{array}\right.
\end{equation}

Then, according to relationships in Table \ref{tab_subsystems_case1}, the probability distribution of the node in level-2 can be describe as
\begin{equation}
\begin{aligned}
U_{1}(z) &=\Psi_{w_{1}}\left(u_{16}^{1}, u_{17}^{1}, u_{18}^{1} u_{19}^{1}\right) \\
&=\sum_{i_{1} 6}^{2} L \sum_{i_{9}=1}^{2} \prod_{j=16}^{19} \mathrm{p}_{j i_{j}} z^{\min \left(g_{16 i_{16}}, L, g_{19 i_{1}}\right)}.
\end{aligned}
\end{equation}

Similarly, probability distributions of other nodes in level-2 are also obtained in the same way. Therefore, the probability distributions of each node in level-2 can be described by $\left\{U_{i} \mid i=1, \cdots, 13\right.$ and $\left.i \neq 3,10\right\}$.

\begin{itemize}
	\item [3)] BN modeling
\end{itemize}

The probability distributions obtained of nodes in level-2 are taken as the input of root nodes, and the BN model is first constructed from level-2 to level-5 as shown in Figure \ref{case1_BN}, where nodes with the same colors are represented in the same level. Then, the Conditional Probability Tables of each node are established. According to the relationships between parent nodes and child nodes from level-2 to level-5, the BN inference is conducted to obtain the probability distributions of nodes 15. Finally, the system reliability can be obtained.

\subsubsection{Result analysis}

\begin{itemize}
	\item [1)] Verification of result accuracy
\end{itemize}

To validate the correctness of the UGF-BN method, BN method is used for comparative analysis, and the results of system reliability inference are shown in Table \ref{tab5}. The UGF-BN method and the BN method are unbiased estimates, so the results of system reliability inference of two method should be the same. According to Table \ref{tab5}, the whole system has four states, where state 4 is the normal working state. Therefore, the system reliability is 0.97001897. Apparently, the results of two methods are exactly the same. the probability values of the two methods are the same for four different states. Therefore, the correctness of the proposed UGF-BN method is validated.


\begin{table}[!htbp]%
	\centering
	\caption{The reliability output comparison between the BN method and the UGF-BN method\label{tab5}}%
	\begin{tabular*}{235pt}{@{\extracolsep\fill}ccc@{\extracolsep\fill}}%
		\toprule
		Method & BN method & UGF-BN method   \\
		\midrule
		State 1 & 0.00020149 & 0.00020149   \\
		State 2 & 0.00986767 & 0.00986767  \\	
		State 3 & 0.01991187 & 0.01991187  \\
		State 4 & 0.97001897 & 0.97001897  \\
		\bottomrule
	\end{tabular*}
\end{table}

\begin{itemize}
	\item [2)] The analysis of time efficiency 
\end{itemize}

\begin{figure}[!htbp]
	\centering
	\includegraphics[scale=0.58]{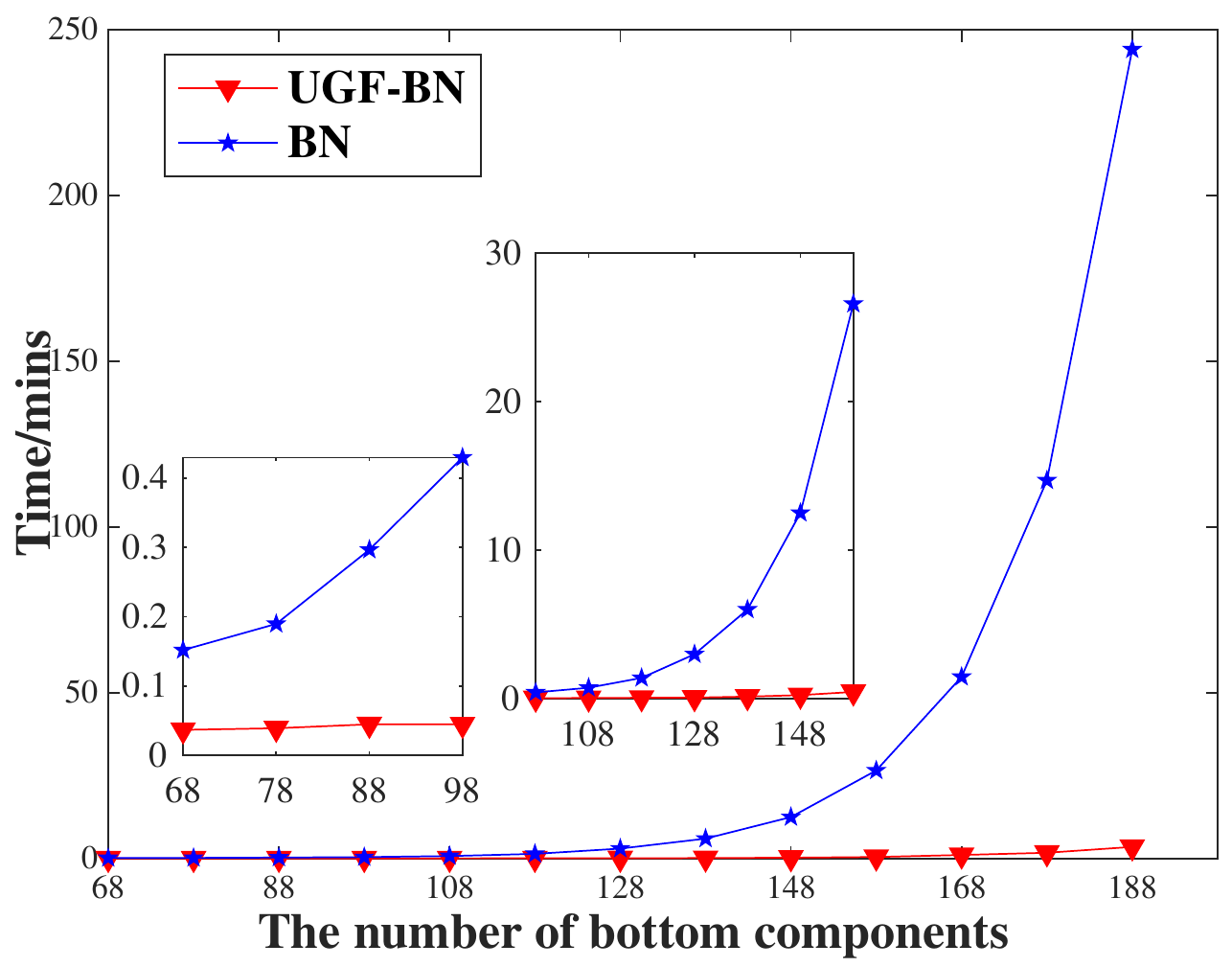}
	\caption{The computational cost of the UGF-BN method and the BN method with the different number of bottom components.}
	\label{case1_time}
\end{figure}

To further verify the efficiency of the UGF-BN method, we design to increase the number of bottom components in level-1. The incremental strategy is to add the bottom components under each node $\mathrm{i} (i=1, \cdots, 13$ but $i \neq 3,10)$. Level-2 has ten nodes, so ten components are added in level-1 at one time. The computational time of the two methods is shown in Figure \ref{case1_time}. According to Figure \ref{case1_time}, when the number of components increases, the computational time curve of the BN method is almost exponentially increased, while the computational time curve of the UGF-BN method increase slowly. Besides, the efficiency advantage of the UGF-BN method is very prominent when the number of components is large. The results show that the UGF-BN method significantly improves the computational time and acquires the system reliability more effectively than the BN method.

\begin{table*}[!htbp]%
	\centering
	\caption{The basic information of different basic units in case 2.	\label{tab_basic_information_case2}}%
	\begin{tabular*}{500pt}{@{\extracolsep\fill}ccc@{\extracolsep\fill}ccc@{\extracolsep\fill}ccc@{\extracolsep\fill}ccc@{\extracolsep\fill}ccc@{\extracolsep\fill}}%
		\toprule
		Component & Mass(Kg) & Power(W) & $Cost(M\$)$ & Min-number& Max-number& $\Psi_{i}$& $\lambda$ \\
		\midrule
		$C_{1}$ & 0.05 &  0.11&	0.1& 1&	9& 3&$1.42 \times 10^{-6}$ \\
		$C_{2}$ & 0.013 & 0.30&	0.2& 2&	8& 1& $1.53 \times 10^{-6}$\\
		$C_{3}$ & 0.014 & 0.40&	0.1& 1&	9& 2& $1.31 \times 10^{-6}$\\
		$C_{4}$ & 0.007 & 0.70&	0.4& 4&	9& 3& $9.83 \times 10^{-7}$\\
		$C_{5}$ & 0.012 & 0.20&	0.6& 3&	8& 1& $1.52 \times 10^{-6}$\\
		$C_{6}$ & 0.025 & 0.50&	0.7& 1&	10& 2& $1.73 \times 10^{-6}$\\
		$C_{7}$ & 0.015 & 0.50&	0.9& 1&	10& 1& $2.13 \times 10^{-6}$\\			
		$C_{8}$ & 0.017 & 0.70& 1.1& 2&	9& 2& $9.53 \times 10^{-7}$\\
		$C_{9}$ & 0.019 & 0.90&	1.3& 2&	9& 1& $9.98 \times 10^{-7}$\\
		$C_{10}$ & 0.020& 0.20&	1.5& 2&	10& 1& $2.13 \times 10^{-6}$\\
		$C_{11}$ & 0.016& 0.60&	1.6& 3&	11& 2& $1.24 \times 10^{-6}$\\
		$C_{12}$ & 0.018& 0.80&	1.7& 4&	11&	1& $1.63 \times 10^{-6}$\\
		$C_{13}$ & 0.020& 0.20&	1.8& 4&	12&	2&$1.45 \times 10^{-6}$\\
		$C_{14}$ & 0.014& 0.40&	1.1& 2&	11&	1& $1.56 \times 10^{-6}$\\
		$C_{15}$ & 0.010& 0.10&	1.0& 1&	10&	2& $1.71 \times 10^{-6}$\\			
		$C_{16}$ & 0.008& 0.80&	0.9& 1&	11&	1& $1.22 \times 10^{-6}$\\
		\bottomrule
	\end{tabular*}
\end{table*}

\begin{figure}[!htbp]
	\centering
	\includegraphics[scale=0.5]{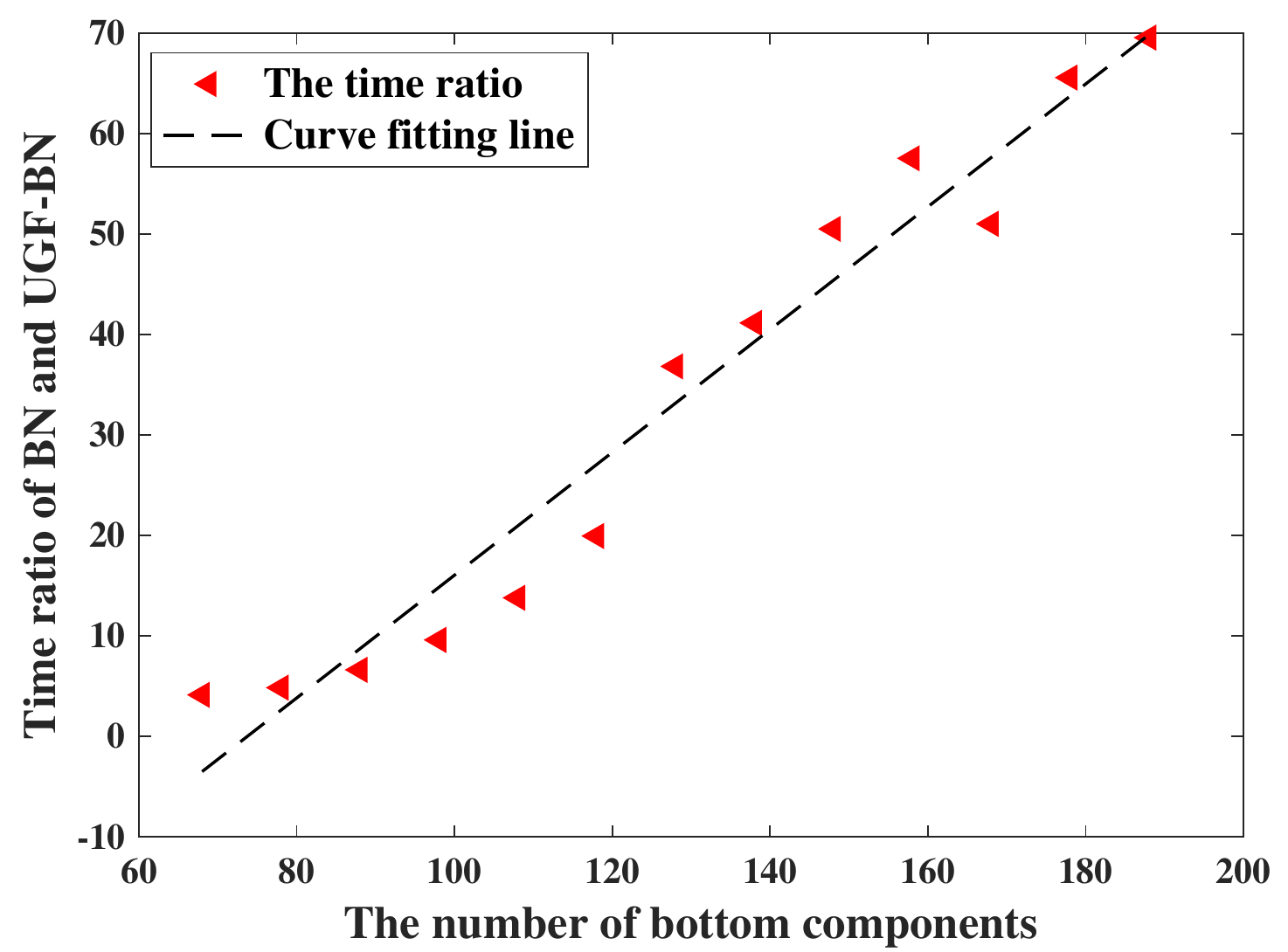}
	\caption{The time ratio of the two methods with the different number of bottom components.}
	\label{case1_time_ratio}
\end{figure}

To further compare the computational cost of the two methods. The time ratio between the BN method and the UGF-BN method is shown in Figure \ref{case1_time_ratio}. According to Figure \ref{case1_time_ratio}, it can be observed that the time ratio of BN method and the UGF-BN method is increasing rapidly, with the increase of the number of bottom components. Especially, we use the linear function curve fit it and it is obvious that the growth trend is almost an equal-proportional line. In conclusion, this case clearly demonstrates that the proposed method is more efficient than the BN method. Moreover, the efficiency advantage of the UGF-BN method becomes more apparent as the number of bottom components increases, which is significant to analyze the complex MSS with an extremely number of bottom components.

\begin{itemize}
	\item [3)] Reliability analysis
\end{itemize}

\begin{figure}[!htbp]
	\centering
	\includegraphics[scale=0.7]{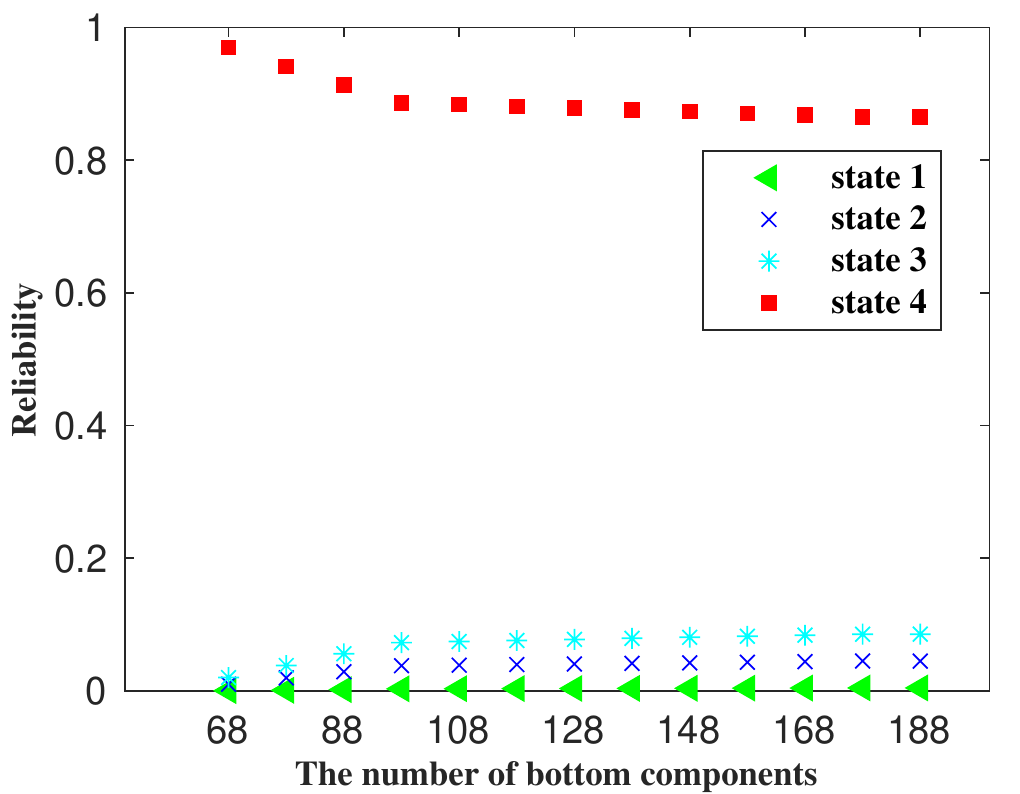}
	\caption{The power driven system reliability with the different number of bottom components.}
	\label{case1_state}
\end{figure}

According to Figure \ref{case1_state}, the system has four states, where state 4 represents normal working, state 1 represents complete failure, state 2 and state 3 denotes degradation. With the increase of the components in level-1, the probability of normal working state decreases gradually, while the probability of other three states keep increasing at a very low level. The trend of probability distribution of the system illustrates that, the probability value of the system at the normal working state is not high with more components in level-1. Therefore, at the beginning of the dynamical driven system design, only two battery packs are used, which can not only improve the reliability of the system, but also reduces the development cost. At the same time, this trend can provide a reference for the system reliability-based design optimization.

\begin{figure*}[!htbp]
	\centering
	\includegraphics[scale=0.28]{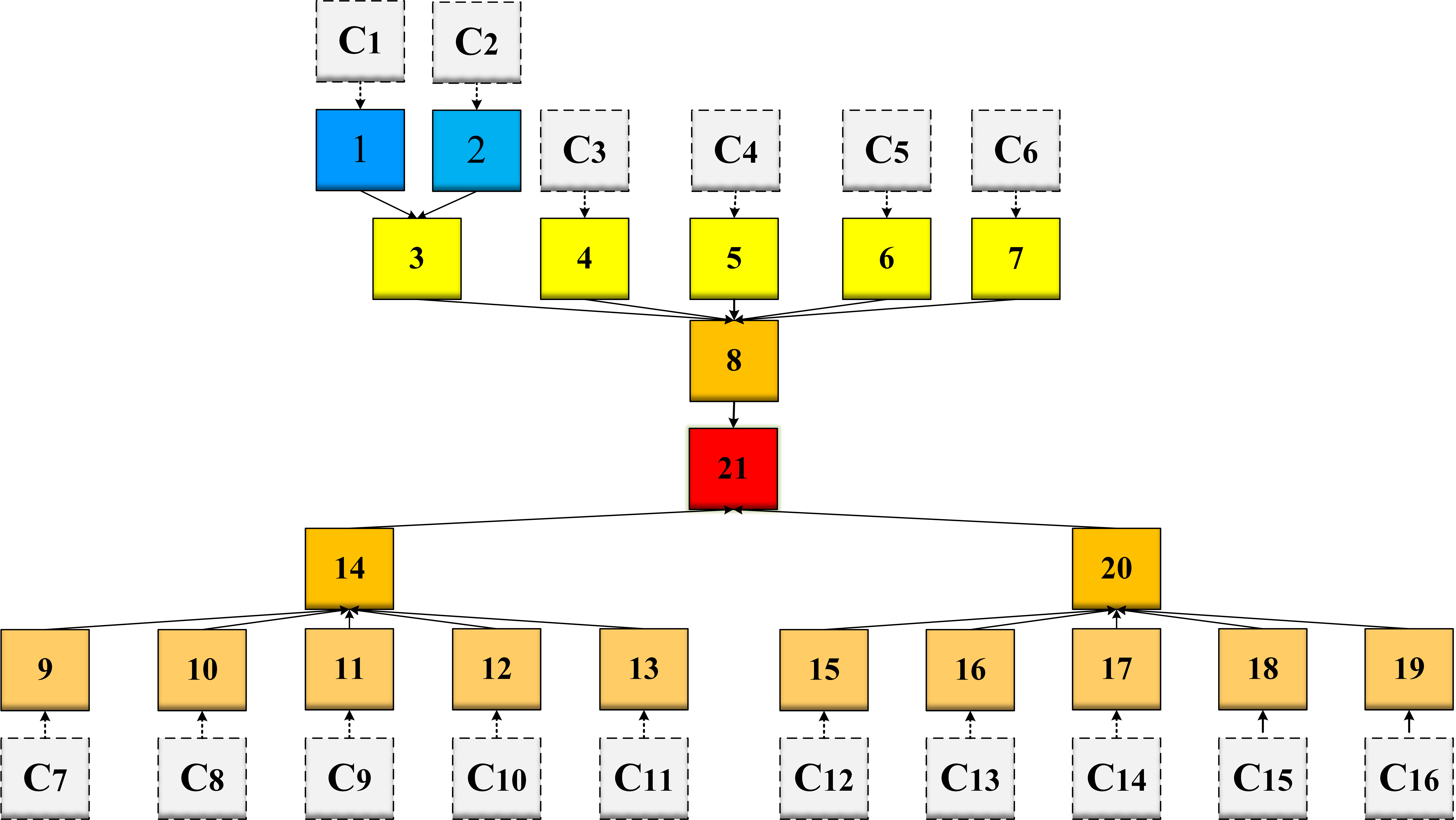}
	\caption{The hierarchical system of the unmanned aerial vehicle dynamical system.}
	\label{case2_hi}
\end{figure*}

\begin{table*}[!htbp]%
	\centering
	\caption{Baseline and optimum design schemes in case 2.	\label{tab_design_schemes_case2}}%
	\begin{tabular*}{280pt}{@{\extracolsep\fill}ccc|@{\extracolsep\fill}ccc@{\extracolsep\fill}ccc@{\extracolsep\fill}}%
		\toprule
		Variable & Baseline & Optimum & Variable & Baseline& Optimum \\
		\midrule
		$C_{1}$ & 10 & 9 & $C_{9}$ & 7 & 2 \\
		$C_{2}$ & 7 & 4 & $C_{10}$ & 12 & 11 \\
		$C_{3}$ & 7 & 2 & $C_{11}$ & 9 & 7 \\
		$C_{4}$ & 9 & 2 & $C_{12}$ & 8 & 5 \\
		$C_{5}$ & 11 & 7 & $C_{13}$ & 10 & 8 \\
		$C_{6}$ & 8 & 2 & $C_{14}$ & 9 & 7 \\
		$C_{7}$ & 12 & 10 & $C_{15}$ & 12 & 11 \\
		$C_{8}$ & 6 & 2 & $C_{16}$ & 8 & 3 \\
		\bottomrule
	\end{tabular*}
\end{table*}

\subsection{Case 2}
\subsubsection{Problem Description}

PX4 is an autopilot software and has five-levels as shown in Figure \ref{case2_hi}. Besides, the PX4 system has four states of the system, which represent states 1 to 4, respectively. The state 4 is the normal working state. Each bottom component is a basic unit $C_{j}(j=1, \cdots, 16)$. And empirical lifetime distributions for basic units $C_{j}$ are exponential distributions, whose parameters are also listed in Table \ref{tab_basic_information_case2}. In Figure \ref{case2_hi}, these pending basic unit ($C_{1}-C_{16}$) are represented as dashed box. The number of basic units $C_{j}$ is designed during the reliability optimization process. The upper and lower bounds of each basic unite $C_{j}$ are $n_{j}^{\max }=14$ and $n_{j}^{\min }=8$, respectively. The budgets of mass, power, development cost and reliability are set to be: $P o_{budget}=150 \mathrm{~W}$, $M_{budget}=15 \mathrm{~kg}$, $D c_{budget}=40 \mathrm{M} \$$, $R_{budget}=0.9$, respectively.

\subsubsection{Reliability modelling}


According to the problem description of the aircraft structure, the hierarchical system is constructed as shown in Figure \ref{case2_hi}, where components with the same color are represented in the same level. For the basic units, once the number is determined, its failure probability in the lifetime t is calculated by
\begin{equation}
F(t)=\int_{0}^{t} f(t) d t=\int_{0}^{t} \lambda e^{-\lambda t} d t=1-e^{-\lambda t}.
\end{equation}
Therefore, the reliability function is $1-F(t)$. 

\begin{figure*}[!htbp]
	\centering
	\includegraphics[scale=0.5]{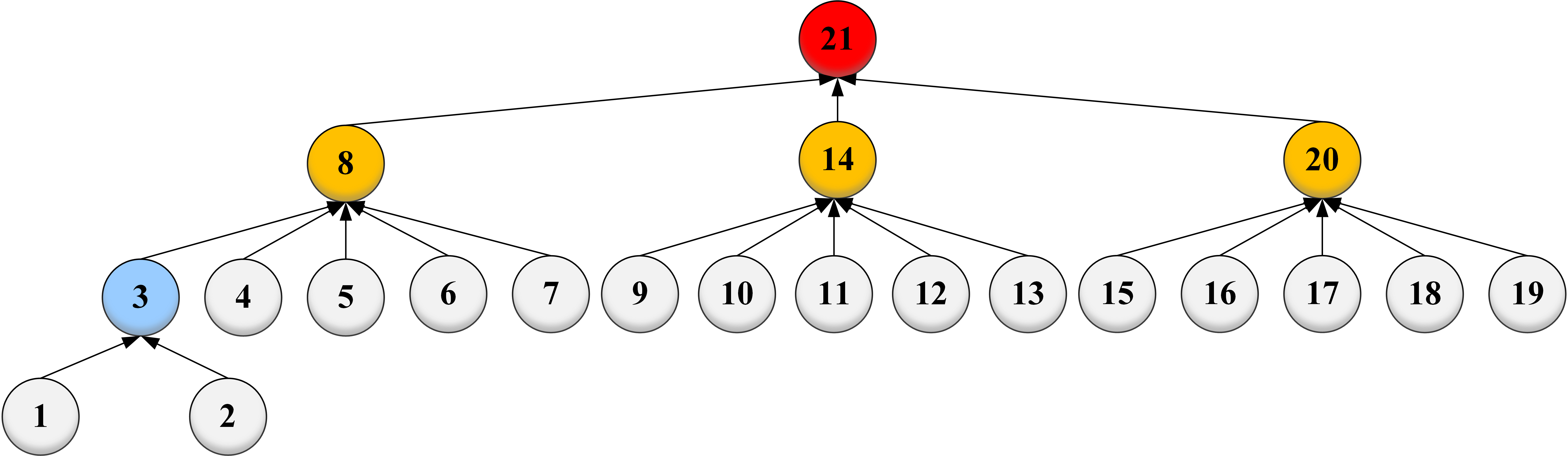}
	\caption{The BN model of PX4.}
	\label{case2_BN}
\end{figure*}

Based on the probability distributions of nodes in level-1, the UGF-BN method is utilized to obtain the system reliability. Firstly, the probability distributions of each node in level-2 are obtained by the UGF method, then BN model is constructed from level-2 to level-5 as shown in Figure \ref{case2_BN}. The probability distributions of each node in level-2 is regarded as the input of the BN model. Finally, the BN method is used to obtain the system reliability.

\begin{table*}[!htbp]%
	\centering
	\caption{The baseline and optimum schemes of mass, power, and cost, and reliability in case 2.\label{tab_mass_case2}}%
    \begin{tabular}{cccccc}
    \hline
Constrains  & Budget & Baseline & Optimum & $\Delta S_{*}^{1}$      & $\Delta S_{*}^{2}$      \\ \hline
Total Mass  & 15     & 17.408   & 9.138   & 16.05\%  & -39.08\% \\
Total Power & 140    & 314.000  & 139.140 & 109.33\% & -7.24\%  \\
Total Cost  & 40     & 53.080   & 32.450  & 32.70\%  & -18.87\% \\
Reliability & 0.9    & 0.745    & 0.913   & -17.20\% & 1.42\%   \\ \hline
    \end{tabular}
\end{table*}

\subsubsection{Optimization modelling}

The mass, power, and cost of each basic units $C_{j}(j=1, \mathrm{K}, 16)$ are given in Table \ref{tab_basic_information_case2} and the total mass, power and cost of models are obtained by Eq.(\ref{eq_sum_mass}), Eq.(\ref{eq_sum_pow}) and Eq.(\ref{eq_sum_dc}). It is noteworthy that the equations for calculating the aircraft mass, power and cost have been simplified in this case to compare different unit number designs schemes estimation, other methods can be found in reference \cite{larson1992space} for different tasks.
The reliability model is calculated by the reliability modeling method based on the UGF-BN method described in the above section. Then the budget of the total mass, the total power and the total cost are set to be $M_{budget}=15 \mathrm{~kg}$, $P o_{budget}=150 W$, $D c_{budget}=40 M \$$ (Design lifetime is 4 years), the aircraft optimization model is described as
\begin{equation}
\left\{\begin{array}{l}
\text { find } n_{j}, j=1, \cdots, 16, \\
\max R_{\text {system}}, \\
\text { s.t. } M_{\text {sum }} \leq 15 \mathrm{~kg}, \\
P o_{\text {sum }} \leq 150 \mathrm{~W}, \\
D c_{\text {sum }} \leq 40 \mathrm{M} \$, \\
8 \leq n_{j} \leq 14.
\end{array}\right.
\end{equation}

Then the Sequence Quadratic Program algorithm is used as a optimization solver, and the stopping criteria are that the size of a step is smaller than $1e-15$ and the maximum iteration number is 500. Finally, optimization results will be obtained.

\subsubsection{Result analysis}
The baseline and optimum schemes of the twelve design variables and constraints are listed in Table \ref{tab_design_schemes_case2} and Table \ref{tab_mass_case2}, respectively. Taking the budget value of mass, power and cost as benchmark, the ratio (denoted as $K_{*}$) and the relative difference (denoted as $\Delta S_{*}$) between the baseline/optimum scheme values and the corresponding budget values can be calculated by:

\begin{equation}
\begin{aligned}
& K_{*}=\frac{\text { baseline }_{*} \text { oroptimum }_{*}}{\text { benchmark }_{*}}, \\
\Delta S_{*}^{1}=& \frac{\text { baseline }_{*}-\text { benchmark }_{*}}{\text { benchmark }_{*}} \%, \\
\Delta S_{*}^{2}=& \frac{\text { optimum }_{*}-\text { benchmark }_{*}}{\text { benchmark }_{*}} \%,
\end{aligned}
\end{equation}
where the subscript $*$ represents mass, cost, or reliability.

\begin{figure}[!h]
	\centering
	\includegraphics[scale=0.62]{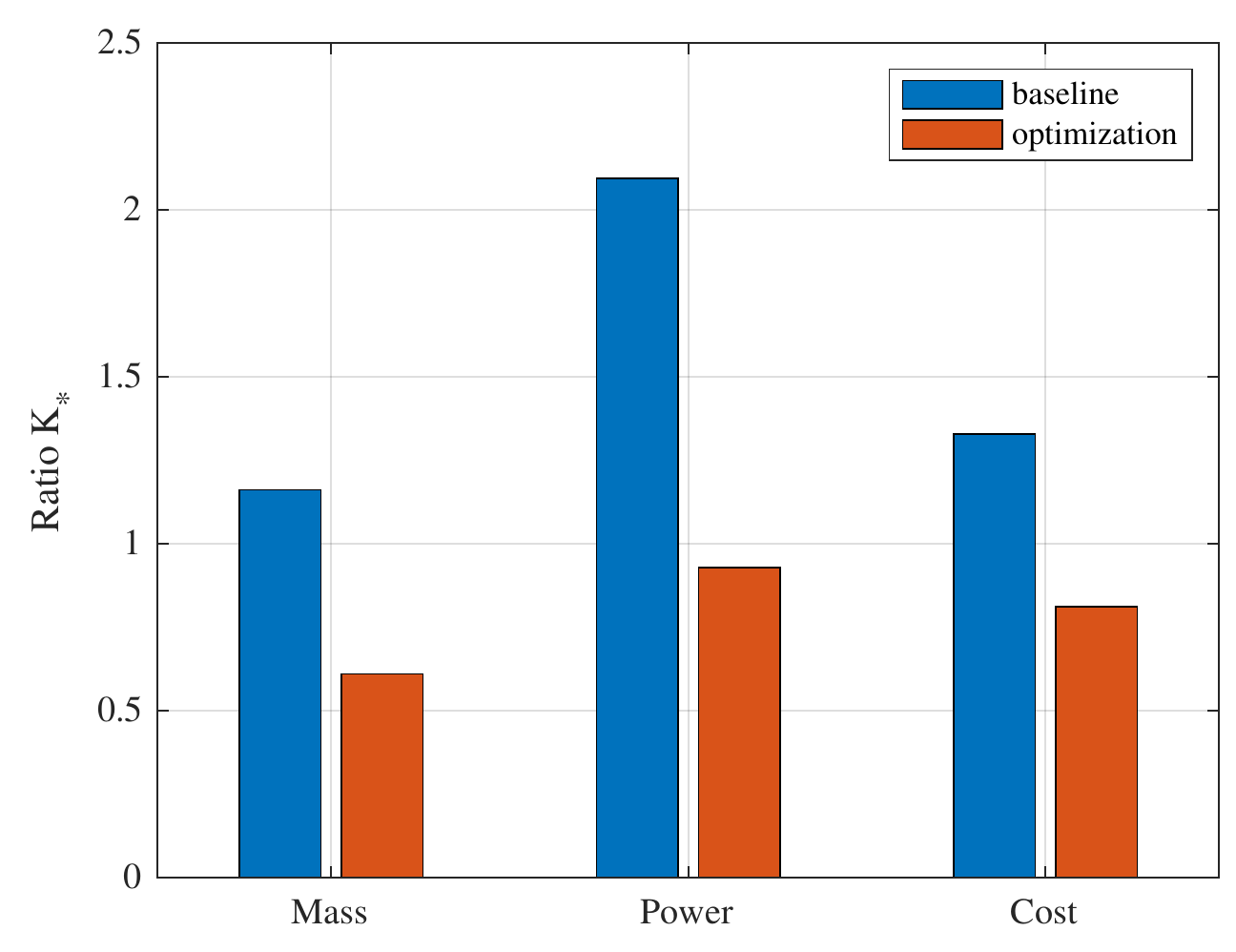}
	\caption{Ratio between the baseline/optimum schemes and the budgets.}
	\label{case2_optium}
\end{figure}

From Table \ref{tab_design_schemes_case2}, it is obvious that the optimum scheme improves a lot, compared with the baseline scheme. From Table \ref{tab_mass_case2}, under the baseline strategy, the total mass and total cost exceed the budget by $16.05\%$ and $32.70\%$, and the reliability is below the budget by $17.20\%$. Especially for the total power, baseline scheme exceeded the budget by $109.33\%$, which means that the baseline scheme greatly violates the constraint. For the optimum scheme, the total mass decreases from 17.408 kg to 9.138 kg, the total power decreases from 314 W to 139.14 W, the total cost decreases from $53.080 M\$$ to $32.450 M\$$ and the reliability improve from 0.745 to 0.913. Besides, as shown in Figure \ref{case2_optium}, the total mass, power and cost are far below the budget as well as the system reliability is much higher than the baseline. Apparently, the optimization effect is remarkable. In other words, the optimization results meet the constraints of the optimization problem. More importantly, the optimization objective of reliability has been greatly improved compared with the baseline.

\hspace{12 pt}
\hspace{12 pt}
\hspace{12 pt}

\begin{figure}[!htbp]
	\centering
	\includegraphics[scale=0.72]{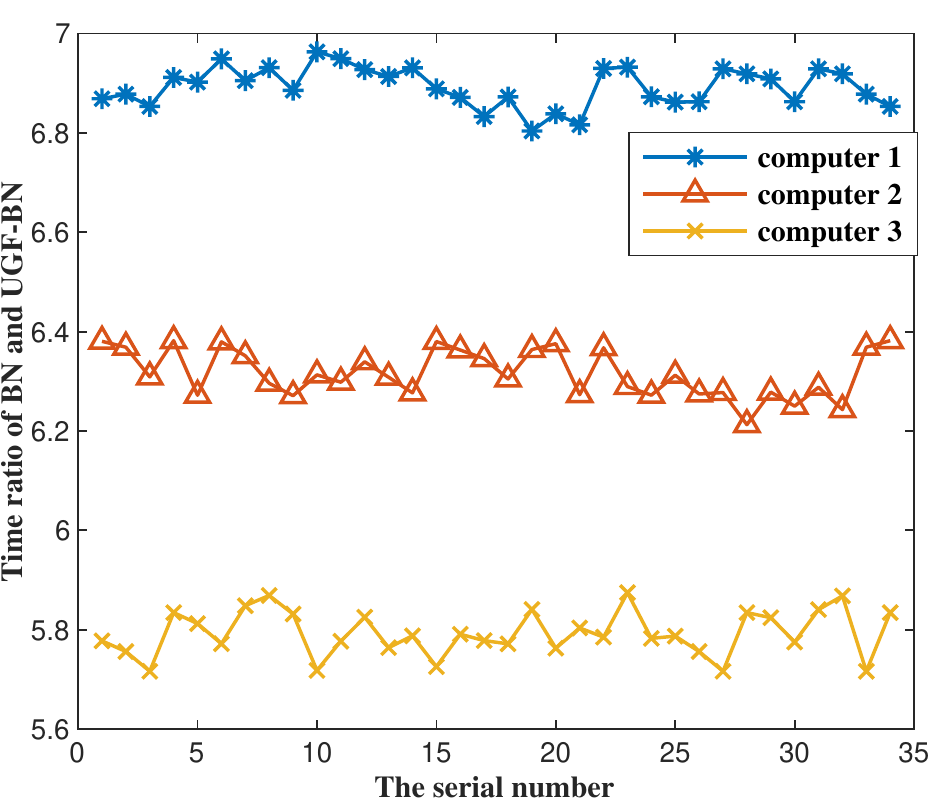}
	\caption{The time ratio of the BN method and the UGF-BN method.}
	\label{case2_time_ratio}
\end{figure}

To further demonstrate the computational efficiency of the UGF-BN method in reliability-based design optimization, the computational time ratio of the BN method and the UGF-BN method for optimization is shown in Figure \ref{case2_time_ratio}. Besides, the time ratio is obtained by testing on three different computers. As shown in the Figure \ref{case2_time_ratio}, the time ratio fluctuates steadily between 5 to 7 times for different computers, which means that computational cost of the UGF-BN method is considerably cheaper than the BN method.

In summary, the optimum scheme significantly reduces the total mass, power and cost. Meanwhile, the optimum scheme greatly improves the system reliability. The aircraft system is a complex MSS with a large number of bottom components, which leads to the exponential increment of time cost. As for the optimum scheme proposed in this paper, the more bottom components, the more obvious the improvement of computational cost. Thus, the optimization design scheme obtained provides a valuable reference for designers.

\vspace{15 pt}

\section{Conclusions}\label{sec6}
In this paper, the reliability modeling of the complex MSS is studied based on the UGF method and the BN method. The UGF method is an important way to obtain the system reliability efficiently by using a fast algebraic procedure, and the BN method has a natural advantage in uncertainty reasoning for the system structure without explicit expressions. The UGF-BN method proposed in this paper combines the respective superiority of two methods. In the UGF-BN method framework, the UGF method is used to analyze the bottom components with a large number, and the BN method is utilized to analyze the subsystem without explicit expressions. The UGF-BN method efficiently obtains the reliability of the complex MSS, which extremely reduces the computational cost. On the above basis, the aircraft reliability-based design optimization model is established by considering the constraints. In the aircraft reliability-based design optimization process, the UGF-BN method greatly reduces the computational cost of the reliability analysis.

Two cases studies are used to demonstrate and validate the proposed method. For case 1, we compare the computational time of the UGF-BN method and the BN method. The UGF-BN method significantly reduces the computational time, proving the efficiency of the proposed method, especially for the complex MSS with an extremely number of bottom components. For case 2, the UGF-BN method is applied to the aircraft reliability-based design optimization to maximize the total reliability with budget constraints on total mass, power and cost. Compared with the baseline scheme, the optimal scheme significantly reduces the total mass, power and cost, and improves the reliability, providing a reference for aircraft reliability design. Besides, the optimization effects of computational time are veriﬁed under three different computers. In the UGF-BN framework, the UGF method is used to analyze the components in level-1, which just provides an idea. Considering that more levels are analyzed by the UGF method, the computational time will further improve in the future study.

\section{Declaration of conflicting interests}
The author(s) declared no potential conflicts of interest
with respect to the research, authorship, and/or publication of this article.

\section*{Funding} 
This work was supported by National Natural Science Foundation of China under Grant No.11725211, 52005505, and 62001502.

\bibliography{reference}%





\end{document}